\documentclass[fleqn,10pt]{wlscirep}
\usepackage[utf8]{inputenc}
\usepackage[T1]{fontenc}
\title{Structuring the local handedness of synthetic chiral light: global chirality versus polarization of chirality}

\author[1,2,*]{Laura Rego}
\author[2,3,+]{David Ayuso}

\affil[1]{Department of Physics, Imperial College London, SW7 2AZ London, United Kingdom}
\affil[2]{Universidad de Salamanca, 37008 Salamanca, Spain}
\affil[3]{Max-Born-Institut, 12489 Berlin, Germany}

\affil[*]{laura.rego@imperial.ac.uk}
\affil[+]{david.ayuso@imperial.ac.uk}

\begin{abstract}
Synthetic chiral light\cite{Ayuso2019NatPhot,Ayuso2021NatComm,Ayuso2022PCCP,Neufeld2021PRR,Katsoulis2021,Mayer2021,Khokhlova2021} enables ultrafast and highly efficient imaging of molecular chirality.
Unlike standard circularly polarized light, the handedness of synthetic chiral light does not rely on the spatial structure of the light field:
it is encoded locally, in the chiral trajectory that the tip of the electric-field vector draws in time, at each point in space.
Being locally chiral, already within the electric-dipole approximation, synthetic chiral light is a highly efficient chiral photonic reagent.
Synthetic chiral light that is locally and globally chiral \cite{Ayuso2019NatPhot} allows us to selectively quench the nonlinear response of a selected molecular enantiomer while maximizing it in its mirror twin at the level of total signal intensities.
Synthetic chiral light that exhibits polarization of chirality\cite{Ayuso2021NatComm} allows us to realize a chiral version of Young's double-slit experiment that leads to enantio-sensitive light bending.
Here we connect these two new concepts and show how one can structure the local and global handedness of synthetic chiral light in space, and how these local and global properties are imprinted in the enantio-sensitive response of the chiral molecules, creating new opportunities for ultrafast, all-optical and highly efficient imaging of molecular chirality.
\end{abstract}
\begin{document}

\flushbottom
\maketitle

\thispagestyle{empty}

\section*{Introduction}

The possibility of structuring the physical properties of light in space at will has opened tremendous possibilities for sculpting light-matter interactions \cite{Andrews2011, Rubinsztein2016, Forbes2021}.
Vortex beams, which exhibit a twisted phase profile and carry orbital angular momentum\cite{Allen1992}, are a typical example of structured light.
Such structured light has found exciting applications across a number of fields, including particle manipulation \cite{Grier2003,Simpson1996}, information transfer \cite{Wang2012}, phase contrast \cite{Furhapter2005} and super-resolution microscopies \cite{Vicidomini2018}, or quantum information \cite{Mair2001}.
Likewise, light beams with structured polarization \cite{Maurer2007, Zhan2009, Beckley2010} have been proven to be unique tools at the nanoscale \cite{Novotny2001, Bautista2012, Wozniak2015} due to their superior focusing properties \cite{Dorn2003}.

In the last decade, structured vortex light has been successfully applied to drive highly non-linear interactions in matter, such as structured high-order harmonic generation (HHG), leading to the creation of ultrashort structured pulses in the extreme-ultraviolet domain\cite{Zurch2012, Hernandez2013, Gariepy2014, Geneaux2016, Kong2017, Hickstein2015, Ellis2018, Azoury2019, Pisanty2019, Delasheras2022,Dorney2019}.
An interesting aspect of this capacity to structure the HHG radiation is that it allows us to spatially separate harmonics radiation with different properties, enabling control over the polarization of ultrashort pulses \cite{Dorney2019, Rego2020} or their spectral content \cite{Rego2022}.

However, it is not necessary to resort to orbital angular momentum to create structured light.
For instance, it is well known that the polarization of a gaussian beam becomes spatially structured upon tight focusing\cite{Bauer2016,Bliokh2015}, or when overlapping two laser pulses that propagate non-collinearly\cite{Bliokh2015,Heyl2014NJP,Pisanty2018NJP}.
Here we exploit the capabilities of structuring light's polarization to structure the local handedness of synthetic chiral light\cite{Ayuso2019NatPhot,Ayuso2021NatComm} in space, and apply such structured locally chiral fields to image chiral molecules on ultrafast timescales.

Chiral molecules exist in pairs of non-superimposable mirror images: the left- and right handed enantiomers, which have identical physical and chemical properties, e.g. melting and boiling points, energy levels, etc., and thus behave identically, unless they interact with something that is also chiral.
Not surprisingly, structured light is finding interesting applications for detecting and even separating opposite molecular enantiomers\cite{Cameron2014JPCA,Cameron2014NJP,Liu2019PCCP}, although these applications are limited by the weakness of non-electric-dipole interactions.
Indeed, the enantio-sensitive response of chiral molecules to an elliptically or circularly polarized fields usually relies on weak magnetic or quadrupole effects which arise beyond the electric-dipole approximation\cite{Barron2004}.
A way around this fundamental limitation is to create a chiral measurement setup \cite{Ordonez2018PRA,Ayuso2022persp},
by recording the photoelectron angular distributions upon ionization with circular
\cite{Ritchie1976PRA,Powis2000JCP,Bowering2001PRL,Garcia2003JCP,Lux2012Angew,Stefan2013JCP,Garcia2013NatComm,Janssen2014PCCP,Lux2015ChemPhysChem,Kastner2016CPC,Comby2016JPCL,Beaulieu2016FD,Beaulieu2017Science,Beaulieu2018NatPhys},
elliptical\cite{Comby2018NatComm},
or two-colour\cite{Demekhin2018PRL,Goetz2019PRL,Rozen2019PRX,Ordonez2022PCCP} driving fields,
measuring the phase of induced nonlinear polarization\cite{Fischer2000PRL,Belkin2001PRL,Fischer2002CPL,Patterson2013},
or creating chiral optical centrifuges\cite{Owens2018PRL,Yachmenev2019PRL,Milner2019PRL}.

Synthetic chiral light\cite{Ayuso2019NatPhot,Ayuso2021NatComm} can be seen as an upgrade with respect to circularly polarized light as a chiral photonic reagent\cite{Ayuso2022persp}.
Such light is locally chiral: the tip of the electric-field vector draws a chiral Lissajous figure in time, in every point in space.
The enantio-sensitive response of chiral molecules to synthetic chiral light is driven by purely electric-dipole interactions, and it is orders to magnitude stronger than with traditional optical fields.
Here we show how to structure the local handedness of synthetic chiral light in space, to imprint the handedness of chiral matter into different macroscopic observables, connecting the recently introduced concepts of global chirality \cite{Ayuso2019NatPhot} and polarization of chirality \cite{Ayuso2021NatComm}.

This paper is structured as follows.
We start by providing a comprehensive analysis of the key aspects of synthetic chiral light.
First, we analyse how the local handedness\cite{Ayuso2019NatPhot} of synthetic chiral light is encoded in the relative phase between its frequency components, and how it can be characterized and quantified using chiral correlation functions\cite{Ayuso2019NatPhot}.
Second, we describe how such a locally chiral field can be generated using a non-collinear laser configuration that allows us to control its local and global properties\cite{Ayuso2019NatPhot,Ayuso2021NatComm}.
We shall see that we can create locally chiral fields which can be (i) perfectly globally chiral\cite{Ayuso2019NatPhot} and chirality unpolaried, (ii) globally achiral but perfectly chirality polarized\cite{Ayuso2021NatComm}, or (iii) something in between, i.e. fields which are both globally chiral and chirality polarized.
Third, we show how these local and global chirality properties are imprinted in the ultrafast nonlinear response of randomly oriented chiral molecules using state-of-the-art computational modelling.
Our analysis provides a simple recipe for creating synthetic chiral light with controlled local and global properties, which allows us to imprint the handedness of randomly oriented chiral molecules into different different macroscopic observables: either the total intensity of HHG, or the direction of harmonic emission.
We conclude by discussing the unique opportunities enabled by such structured locally chiral fields.

\section*{The local handedness of synthetic chiral light}

\begin{figure}[ht]
\centering
\includegraphics[width=\linewidth, keepaspectratio=true]{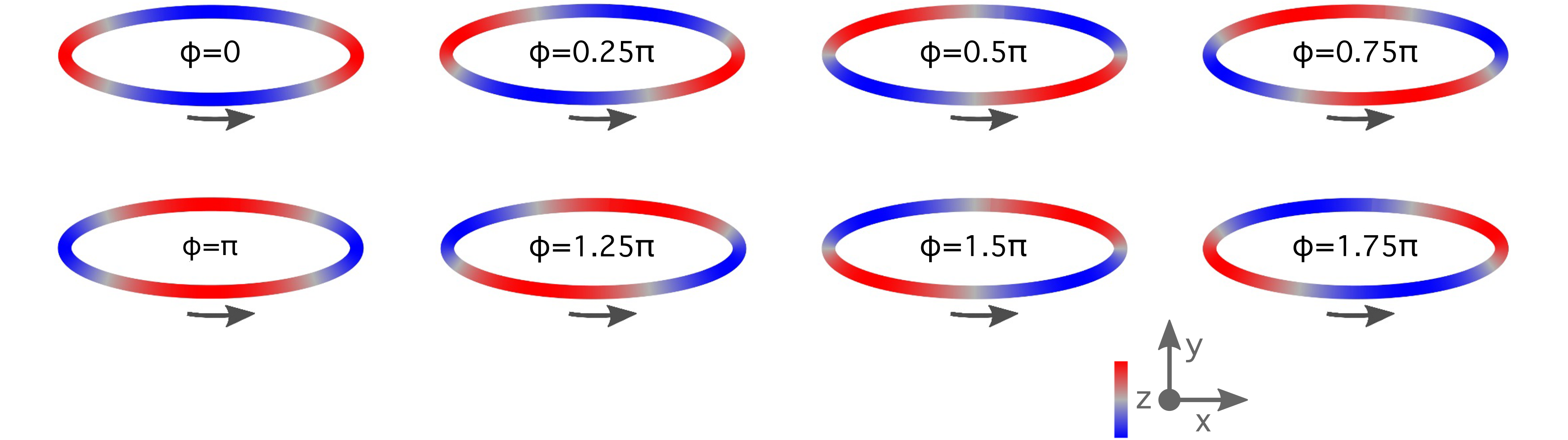}
\caption{\textbf{Local chirality.} Lissajous figures of the locally chiral field of Eq. \ref{eq_Ew2w} for $\varepsilon>0$, $E_{2\omega}>0$, and for different values of the two-colour phase delay $\phi$.
Red and blue indicate positive and negative values of $E_z$, respectively. 
Note that changing the the two-colour phase delay by $\pi$ reverses the field's handedness.
The upper and lower figures in the same column represent field enantiomers.}
\label{fig_Lissajous}
\end{figure}

A field is locally chiral if the Lissajous figure characterizing the temporal evolution of the electric-field vector is a chiral structure\cite{Ayuso2019NatPhot} which cannot be superimposed to its mirror image.
To fulfil this requirement, the field needs to be three-dimensional and contain two \cite{Ayuso2019NatPhot,Ayuso2021NatComm,Ayuso2022PCCP,Neufeld2021PRR,Katsoulis2021,Mayer2021} or more\cite{Khokhlova2021,Kral2003PRL,Gerbasi2004JCP,Eibenberger2017PRL,Shubert2016,Perez2017} frequency components.
Here we use the locally chiral field introduced in \cite{Ayuso2019NatPhot}, where the electric-field vector is elliptically polarized at frequency $\omega$ in the $xy$ plane and it has a $2\omega$ frequency component along $z$,
\begin{equation}\label{eq_Ew2w}
\textbf{E}(t) = E_{\omega} \big[ \cos(\omega t) \hat{\bold{x}} + \varepsilon\sin(\omega t) \hat{\bold{y}} \big] + E_{2\omega} \cos(2\omega t + \phi) \hat{\bold{z}}
\end{equation}
Fig. \ref{fig_Lissajous} shows the Lissajous figure that the tip of the electric field vector $\textbf{E}$ draws in time, for different values of the two-colour phase delay $\phi$.
These Lissajous figures are chiral ``objects'' which cannot be superimposed to their mirror images by rotation.
Their handedness depends on the two-colour phase delay $\phi$ and on the sign of the ellipticity $\varepsilon$.
Note that changing $\phi$ by $\pi$ is equivalent to reflecting the field on the $xy$ plane, and thus this operation reverses the field's handedness.
This means that the fields with, e.g. $\phi=0$ (upper left Lissajous figure in Fig. \ref{fig_Lissajous}) and with $\phi=\pi$ (lower left figure) are enantiomers.
Likewise, changing the sign of $\varepsilon$ is equivalent to reflection on the $xz$ plane (up to rotation on the $x$ axis) and it also reverses the field's chirality.

There is a key fundamental difference between the (local) handedness of synthetic chiral light and the (non local) handedness of circularly polarized light:
a circularly polarized wave can either be left- or right-handed,
whereas the chirality our locally chiral field depends on the two-colour phase delay $\phi$ (see Fig. \ref{fig_Lissajous}), which is a continuous and controllable variable.
This means that synthetic chiral light can have an infinite set of different handedness, which can be controlled by controlling $\phi$, creating powerful opportunities for controlling the enantio-sensitive nonlinear response of chiral matter.

\subsection*{Chiral correlation functions and control over the enantio-sensitive response of chiral matter}

\begin{figure}[ht]
\centering
\includegraphics[width=12cm, keepaspectratio=true]{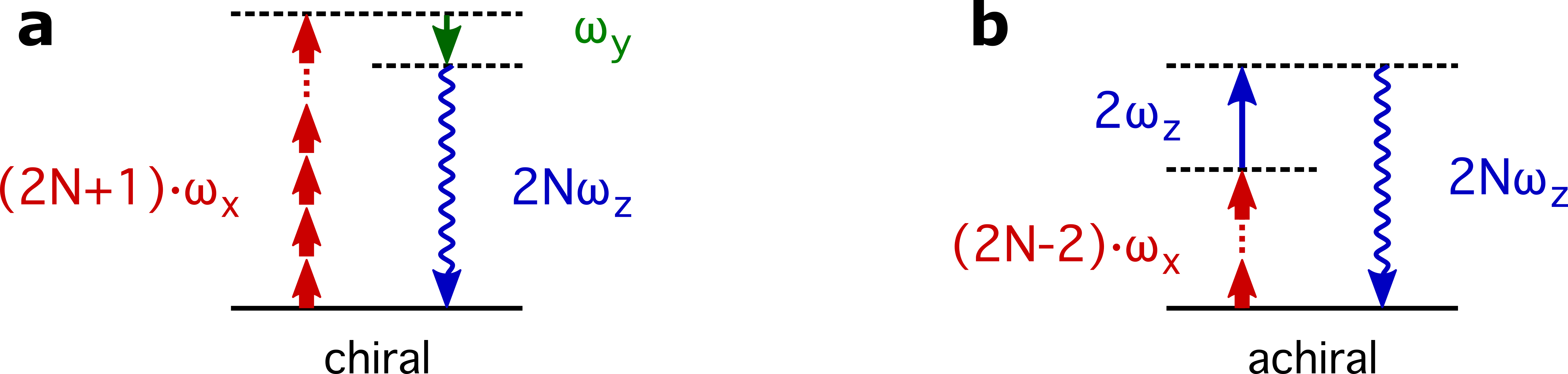}
\caption{\textbf{Multi-photon diagrams} describing the enantio-sensitive response of isotropic chiral matter to synthetic chiral light in the perturbative regime, see main text.
 }\label{fig_diagram}
\end{figure}

The enantio-sensitive response of chiral matter to synthetic chiral light is nonlinear, and it relies on the interference between two contributions to light-induced polarization \cite{Ayuso2019NatPhot}.
The corresponding multiphoton diagrams are shown in Fig. \ref{fig_diagram}.
Here we assume $\varepsilon<<1$ and $E_{2\omega}<<E_{\omega}$, so the molecules can absorb several photons from $E_x$, but their response to $E_y$ and $E_z$ is linear.
The chiral pathway (Fig. \ref{fig_diagram}a) involves absorption of $2N+1$ photons from $E_x$ and emission of $1$ photon into $E_y$.
This pathway is exclusive of chiral media and it leads to polarization at frequency $2N\omega$ along $z$.
The induced polarization is exactly out of phase in molecular opposite enantiomers.
In the achiral pathway (Fig. \ref{fig_diagram}b), which is not sensitive to the medium's handedness, the molecules absorb $2N-2$ photons from $E_x$ and $1$ photon from $E_z$, also leading to polarization at frequency $2N\omega$ along $z$.
We can adjust the relative amplitude and phase of the achiral contribution by controlling $E_{2\omega}$ and $\phi$, and thus maximize the nonlinear response at frequency $2N\omega$ in a selected molecular enantiomer while fully quenching it in its mirror twin.

The local handedness of synthetic chiral light can be quantified using chiral correlation functions\cite{Ayuso2019NatPhot}.
For the locally chiral field considered here, the lowest-order chiral correlation function is $h^{(5)}$, which can be written in the frequency domain as\cite{Ayuso2019NatPhot}
\begin{equation}\label{eq_h5}
h^{(5)} = \big\{ \textbf{F}^*(2\omega) \cdot [\textbf{F}^*(\omega) \times \textbf{F}(\omega)] \big\} [\textbf{F}^*(\omega) \cdot \textbf{F}(\omega)] 
= 2i \varepsilon E_{\omega}^4 E_{2\omega} (1-\varepsilon^2) e^{i\phi}
\end{equation}
where $\textbf{F}$ is the Fourier component of the locally chiral field (Eq. \ref{eq_Ew2w}).
The field's chirality is encoded in the phase of $h^{(5)}$, which depends on the two-colour phase delay $\phi$ and on the sign of $\varepsilon$ and $E_{2\omega}$.

Chiral correlation functions characterize the local handedness of synthetic chiral light, but also the enantio-sensitive response of chiral matter.
In particular, $h^{(5)}$ describes the interference between the two pathways giving rise to enantio-sensitive polarization at frequency $2\omega$ (Fig. \ref{fig_diagram} with $N=1$).
The enantio-sensitive response at higher-order (even) harmonic frequencies can be controlled and characterized via higher-order chiral correlation functions.
For the locally chiral field considered here, assuming $\varepsilon<<1$ and $E_{2\omega}<<E_{\omega}$, we have:
 \begin{equation}
h^{(2N+1)} = 2i \varepsilon E_{\omega}^{2N} E_{2\omega} e^{i\phi}
\end{equation}
Note that the phase of $h^{(2N+1)}$ depends on $\phi$ and on the signs of $\varepsilon$ and $E_{2\omega}$ for all $N$.
In general $h^{(2N+1)}$, quantifies the interference between the achiral and chiral contributions to light induced polarization at frequency $2N\omega$ (Fig. \ref{fig_diagram}).

\section*{Structuring light's local handedness in space}

\begin{figure}[ht]
\centering
\includegraphics[width=\linewidth, keepaspectratio=true]{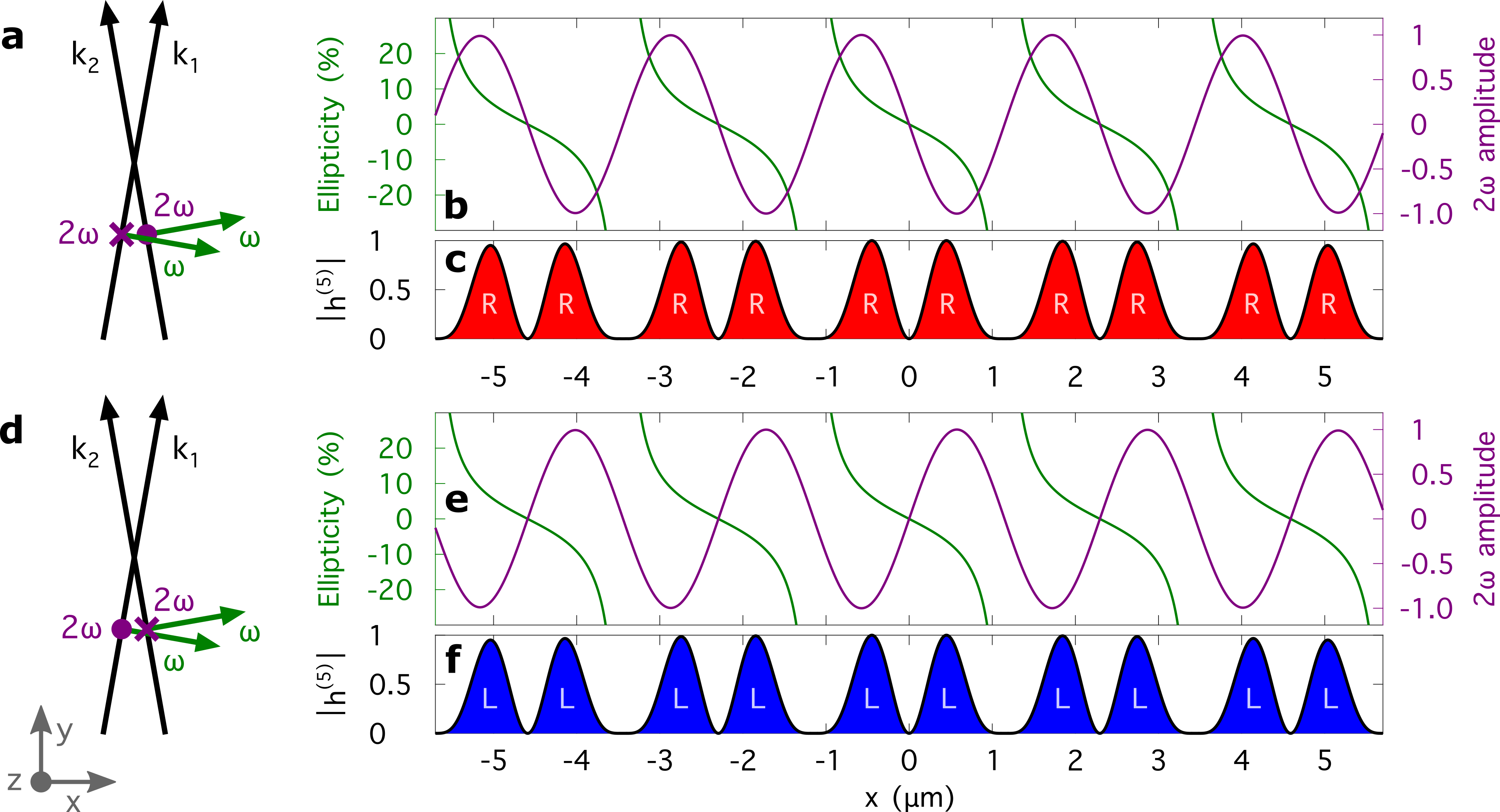}
\caption{\textbf{Synthetic chiral light that is globally chiral.}
\textbf{a,} Schematic representation of the non-collinear setup to create synthetic chiral light that maintains its handedness in space: the two beams are linearly polarized at frequency $\omega$ in the $xy$ plane of propagation, and have a $2\omega$ frequency component orthogonal to this plane, with opposite two-colour phase delay in the two beams $\phi_1=\phi_2+\pi$.
\textbf{b,} Forward ellipticity in the $\omega$-field component (green curve) and normalized $2\omega$-field amplitude (purple curve) along the transverse coordinate $x$, for $\phi_1=\phi_2=0$.
These ellipticity and amplitude gratings change sign exactly at the same positions.
\textbf{c,} Amplitude (black line) of and phase (colour) of $h^{(5)}$, which characterizes the field's local chirality.
The red colour indicates $\arg(h^{(5)})=\pi/2$.
Note that the field's handedness is maintained in space.
\textbf{d-f,} Changing the $2\omega$ phase delays in the two beams by $\pi$ (\textbf{d}) changes the sign of $E_{2\omega}$ in each point in space (\textbf{e}), reversing the field's local handedness (\textbf{f}) while keeping the overall structure globally chiral.
The blue colour in \textbf{f} in indicates $\arg(h^{(5)})=-\pi/2$.
}
\label{fig_setup_g}
\end{figure}

The locally chiral field presented in the previous section can be created\cite{Ayuso2019NatPhot,Ayuso2021NatComm} using two laser beams that propagate non-collinearly in the $xy$ plane, at small angles $\pm\alpha$ with respect to the $y$ axis, see Fig. \ref{fig_setup_g}a.
Each beam carries two cross-polarized colours: the fundamental $\omega$ frequency, polarized the $xy$ propagation plane, and its second harmonic, polarized along $z$.
The electric-field vector in each individual laser beam ($n=1,2$) can be written as  
\begin{equation}\label{eq_En}
\textbf{E}_n(\textbf{r},t) = E^{(0)}_{\omega} \Re\big\{ f_{n,\omega}(\textbf{r},t) e^{\textbf{k}_n\cdot\textbf{r} -  \omega t} \big\} \hat{\textbf{e}}_n
+ E^{(0)}_{2\omega} \Re\big\{ f_{n,2\omega}(\textbf{r},t) e^{2\textbf{k}_n\cdot\textbf{r} - 2\omega t - \phi_{n} } \big\} \hat{\textbf{z}}
\end{equation}
where $E^{(0)}_{\omega}$ and $E^{(0)}_{2\omega}$ are the electric-field amplitudes, $f_{n,\omega}$ and $f_{n,2\omega}$ are envelope functions describing the temporal evolution of the fields and their spatial Gaussian profiles\cite{book_Boyd_NLO}, 
$\textbf{k}_{1,2} = \pm k \sin(\alpha) \hat{\textbf{x}} + k \cos(\alpha) \hat{\textbf{y}}$,
with $k=2\pi/\lambda$ and $\lambda$ being the fundamental wavelength,
$\phi_{n}$ is the two-colour phase delay in each beam,
and $\hat{\textbf{e}}_{1,2} = \cos(\alpha) \hat{\textbf{x}} \mp \sin(\alpha) \hat{\textbf{y}}$.

For $\alpha\neq0$, the beams propagate non-collinearly, and thus the $\hat{\textbf{e}}_1\neq\hat{\textbf{e}}_2$.
That is, the $\omega$-field component of the electric-field vector in the first beam is not parallel to the $\omega$-field component in the second beam.
As a result, in the overlap region, the total electric-field vector becomes elliptically polarized at frequency $\omega$ in the $xy$ plane, with the minor ellipticity component along the $y$ direction (we assume small $\alpha$).
The combination of this elliptical polarization with the linearly polarized component at frequency $2\omega$ creates three-dimensional chiral Lissajous figures like the ones presented in Fig. \ref{fig_Lissajous}.
Note that, for $\alpha\neq0$, the projection of $\textbf{k}_n$ over the $x$ axis is different in each beam, $\textbf{x}\cdot\textbf{k}_1\neq \textbf{x}\cdot\textbf{k}_2$, and thus the relative phase between the frequency components in different beams changes along $x$, creating amplitude and ellipticity gratings in this direction.
As we show in the following, control over such gratings enables control over the global properties of the locally chiral field, and thus over the enantio-sensitive macroscopic response of chiral molecules.

We have modelled the highly nonlinear interaction of synthetic chiral light with randomly oriented chiral molecules in thin media, such as flat liquid microjets \cite{Galinis2017,Luu2018}, where $\Delta y$ is small, or in the gas phase \cite{Cireasa2015NatPhys}, where the two frequency components of the driving field can propagate with approximately the same velocity.
Let us set, for simplicity, $y=z=0$, and write the total electric field resulting from adding the contributions from the two individual beams (Eq. \ref{eq_En}) as
\begin{equation}\label{eq_E_xt}
\textbf{E}(x,t) = E_{\omega}(x,t) \big[ \cos(\omega t) \hat{\bold{x}} + \varepsilon \sin(\omega t) \hat{\bold{y}} \big] + E_{2\omega}(x,t) \cos(2\omega t + \phi_+) \hat{\bold{z}},
\end{equation}
with
\begin{align}
E_{\omega}(x,t)  &= 2 E^{(0)}_{\omega} A_{\omega}(t) e^{-x^2/w^2} \cos(\alpha) \cos( k_\alpha x), \\
E_{2\omega}(x,t) &= 2 E^{(0)}_{2\omega} A_{2\omega}(t) e^{-x^2/w^2}              \cos(2k_\alpha x + \phi_-) \label{eq_E2w}, \\
\varepsilon(x) &= - \tan(\alpha) \tan(k_\alpha x) \label{eq_eps},
\end{align}
where $A_{\omega}$ and $A_{2\omega}$ are envelope functions, $w$ is the beam waist, $k_\alpha = k \sin(\alpha)$, and
\begin{equation}\label{eq_phi_pm}
\phi_{\pm} = \frac{\phi_{2}\pm\phi_{1}}{2}.
\end{equation}
We have neglected the effect of the Gouy phase and the wavefront curvature.

We emphasise that the two laser beams are, individually, not locally chiral.
The shape of the Lissajous figure that the tip of the electric field vector draws in time, in each individual beam, depends on the two-colour phase delay $\phi_{n}$, but it is always confined to a two-dimensional plane orthogonal to $\textbf{k}_n$, and therefore achiral.
However, because the two beams propagate non-collinearly ($\alpha\neq0$), the total electric-field vector resulting from combining the two beams becomes three-dimensional and chiral, yielding chiral Lissajous figures like the ones shown in Fig. \ref{fig_Lissajous}.
Indeed, the field in Eq. \ref{eq_E_xt} is elliptically polarized in the $xy$ plane at frequency $\omega$, with the main polarization component along $x$ and the minor component along the propagation direction $y$, and it has a $2\omega$ frequency component along $z$.
Note that this ``forward'' ellipticity $\varepsilon$ is an essential aspect of this configuration, as it enables the Lissajous figure to be three-dimensional and chiral.
Its strength is proportional the opening angle between the two beams for small values of $\alpha$ (see Eq. \ref{eq_eps}), and it vanishes if the beams are collinear.
Note also that the electric-field amplitudes $E_\omega$ and $E_{2\omega}$, as well as $\varepsilon$, change periodically along $x$, and that the period of these spatial oscillations is proportional to $1/\alpha$, for small $\alpha$.

The averaged sum $\phi_+$ and difference $\phi_-$ phase delays defined in Eq. \ref{eq_phi_pm} allow us to simplify the analysis of the field properties.
Loosely speaking, $\phi_+$ controls the shape of the 3D Lissajous figure, see Eq. \ref{eq_E_xt}.
It determines whether the projection of the Lissajous figure over the $xz$ plane has the shape of an infinite symbol $\infty$, the shape of a smile $\smile$, or something in between, see Fig. \ref{fig_diagram}.
Note that this shape is maintained in space.
That is, if our field is $\smile$-shaped in one point, it will be $\smile$-shaped everywhere in space.
However, its orientation in with respect to the elliptical component may change, because the signs of $E_{2\omega}$ and $\varepsilon$ change as we move along $x$ (see Eqs. \ref{eq_E2w} and \ref{eq_eps}).
If the relative orientation of the $\smile$-shaped component of the Lissajous figure with respect to its elliptical component is the same everywhere in space, then the field's handedness will be maintained.
However, if the relative orientation between them changes, the field's chirality will be reversed periodically in space.
The value of $\phi_-$ determines whether this relative orientation, and thus the field's local chirality, changes periodically in space or is maintained globally, see Eqs. \ref{eq_E_xt} and \ref{eq_E2w}.

As we describe in the following, by controlling $\phi_-$ we can create synthetic chiral light that is globally chiral or globally achiral, which may or may not be chirality polarized. 

\subsection*{Global chirality}

We now analyze how the local handedness of our field (Eq. \ref{eq_E_xt}), which can be characterized by its lowest-order chiral correlation function $h^{(5)}$ (Eq. \ref{eq_h5}), is structured in space.
There are two $x$-dependent terms in Eq. \ref{eq_E_xt} which can modify the field's handedness: $E_{2\omega}$ (Eq. \ref{eq_E2w}) and $\varepsilon$ (Eq. \ref{eq_eps}).
They change sign along $x$ with the same spatial periodicity, and the relative phase between these gratings is controlled by the difference phase delay $\phi_-$.
The key to engineer synthetic chiral light that maintains the same handedness in space is to impose that $\varepsilon$ and $E_{2\omega}$ change sign exactly at the same positions.
Eqs. \ref{eq_E_xt} and \ref{eq_E2w} show that this requirement is fulfilled if $\phi_-=\pi/2$. i.e. if the beams have opposite two-colour phase delay: $\phi_2=\phi_1+\pi$.
Then, by modifying $\phi_+$ while keeping $\phi_-=\pi/2$ constant, i.e. by adjusting $\phi_1$ and $\phi_2$ synchronously while keeping $\phi_2=\phi_1+\pi$, we can tailor the shape of the chiral Lissajous figure (see Fig. \ref{fig_Lissajous}) in a way that it has exactly the same chirality everywhere in space.
Fig. \ref{fig_setup_g} shows how, by imposing this condition, $\varepsilon$ and $E_{2\omega}$ change sign exactly at the same positions (Fig. \ref{fig_setup_g}b).
As a result, the field's local handedness, characterized by the phase of $h^{(5)}$, is maintained in space, see Fig. \ref{fig_setup_g}c.
If we change the $2\omega$ phase delays in the two beams by $\pi$ (Fig. \ref{fig_setup_g}a-c versus Fig. \ref{fig_setup_g}d-f),
$E_{2\omega}$ changes sign at every point in space (Fig. \ref{fig_setup_g}e), and thus the field's local handedness is reversed globally (Fig. \ref{fig_setup_g}f), in a way that the overall structure remains globally chiral.
For illustration purposes, we call the field with $\arg\{h^{(5)}\}=-0.5\pi$ left-handed, and the field with $\arg\{h^{(5)}\}=0.5\pi$ right-handed, although we note this is a completely arbitrary choice.
As above described, while we can always identify pairs of field enantiomers, the local handedness of synthetic chiral light, characterized by the phase of $h^{(5)}$, is a continuous variable.

The global handedness of synthetic chiral light $h_0$ can be quantified by integrating $h^{(5)}$ in space\cite{Ayuso2019NatPhot},
\begin{equation}\label{eq_h0}
h_0 = \int h^{(5)}(x) dx
\end{equation}
The amplitude of $h_0$ maximizes when $\phi_2=\phi_2\pm\pi$ because, in these situations, the field has the same local chirality $h^{(5)}$ everywhere in space, as shown in Fig. \ref{fig_setup_g}.

\begin{figure}[ht]
\centering
\includegraphics[width=\linewidth, keepaspectratio=true]{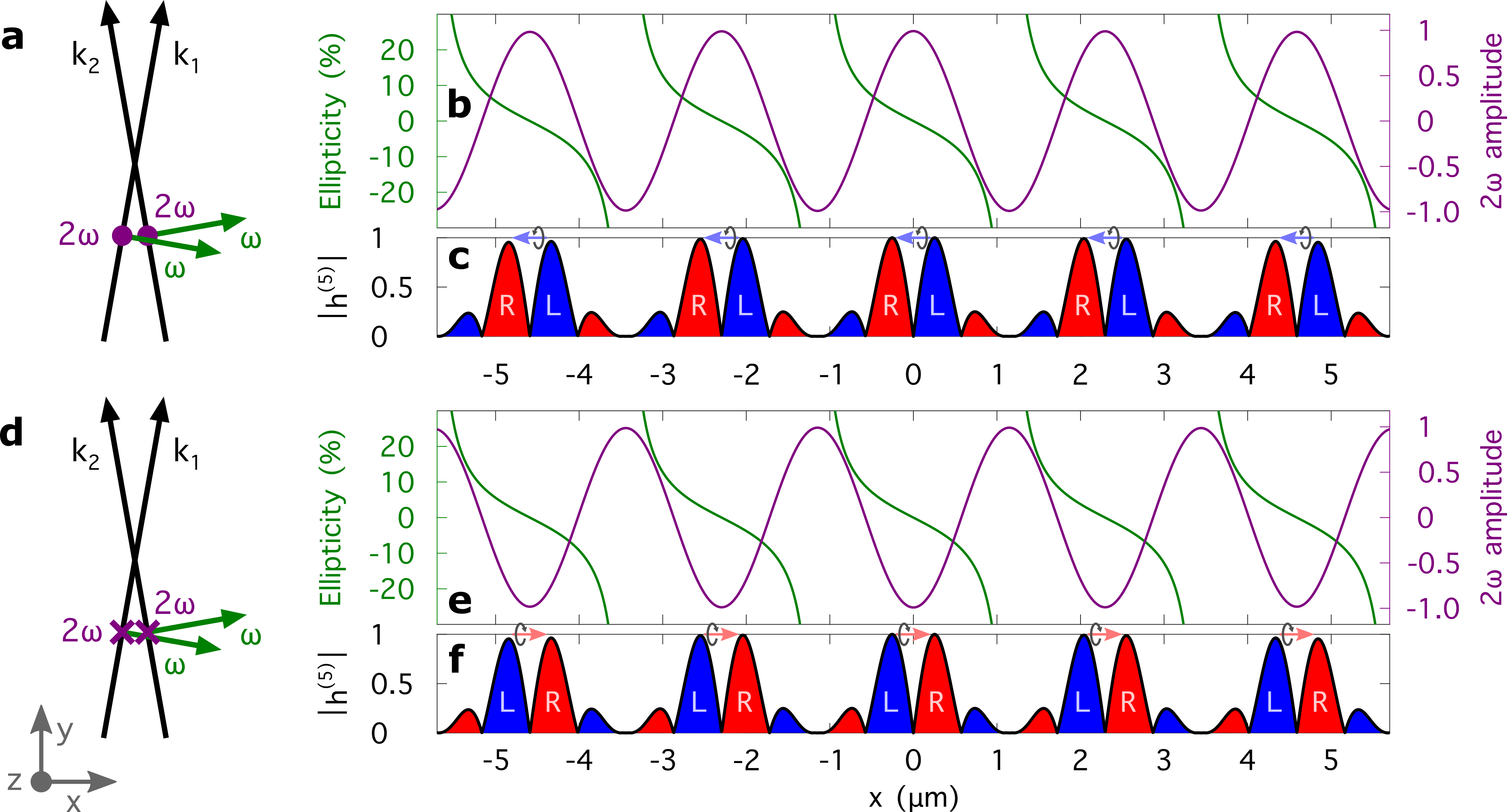}
\caption{\textbf{Synthetic chiral light that is globally achiral, but chirality polarized.}
\textbf{a,} Schematic representation of the non-collinear setup to create synthetic chiral light that changes handedness periodically in space and it has the same amount of opposite field enantiomers: the two beams are linearly polarized at frequency $\omega$ in the $xy$ plane of propagation, and have a $2\omega$ frequency component orthogonal to this plane, with the same two-colour phase delay in the two beams $\phi_1=\phi_2$.
\textbf{b,} Forward ellipticity in the $\omega$-field component (green curve) and normalized $2\omega$-field amplitude (purple curve) along the transverse coordinate $x$, for $\phi_1=0$ and $\phi_2=\pi$.
These ellipticity and amplitude gratings change sign at different positions.
\textbf{c,} Amplitude (black line) of and phase (colour) of $h^{(5)}$, which characterizes the field's local chirality.
The red colour indicates $\arg(h^{(5)})=\pi/2$ and blue colour indicates $\arg(h^{(5)})=-\pi/2$.
In this configuration, the field changes chirality periodically in space, creating dipoles of chirality, and the overall field has polarization of chirality.
\textbf{d-f,} Changing the $2\omega$ phase delays in the two beams by $\pi$ (\textbf{d}) changes the sign of $E_{2\omega}$ in each point in space (\textbf{e}), reversing the field's local handedness (\textbf{f}).
As a result, the field remains globally achiral and chirality polarized, and the direction of polarization of chirality changes direction.}
\label{fig_setup_p}
\end{figure}

\subsection*{Polarization of chirality}

Naively, one could think synthetic chiral light that is not globally chiral could not be used to identify the chirality of matter, but this is not necessarily the case \cite{Ayuso2021NatComm}.
If we set $\phi_1=\phi_2$, see Fig. \ref{fig_setup_p}a, then $\varepsilon$ and $E_{2\omega}$ change sign in different points in space, with a spatial phase delay of $\pi/2$, as shown Fig. \ref{fig_setup_p}b for $\phi_1=\phi_2=0$.
In this case, the field's local chirality changes periodically in space, as shown in as shown Fig. \ref{fig_setup_p}c, and the overall structure is globally achiral, $h_0=0$.
Indeed, the field has mirror symmetry with respect to e.g. the $xz$ plane (up to a global temporal delay).

Interestingly, while the field presented in Fig. \ref{fig_setup_p}a-c is not globally chiral, its handedness is spatially structured, and we find dipoles of synthetic chiral light\cite{Ayuso2021NatComm}: pairs of fields with opposite handedness, see  Fig. \ref{fig_setup_p}c, in analogy with a dipole of electric charge.
The dipole of chirality is a pseudo-vector that points from the left-handed field to the right-handed field\cite{Ayuso2019NatPhot} (Fig. \ref{fig_setup_p}c).  
Crucially, these dipoles of chirality are oriented: here,  the right-handed field is always on the left, whereas the left-handed field is always on the right.
As a result, we find the following distribution of handedness as we move towards positive values of $x$: ... RL RL RL..., and the overall structure has polarization of chirality\cite{Ayuso2021NatComm}, in analogy with the polarization of electric charge in a dielectric.

The amplitude and direction of polarization of chirality can be controlled by controlling the two-colour phase delays in our setup, see Fig. 4.
If we change the phase delay in the two beams by $\pi$ (Fig. \ref{fig_setup_p}a-c versus Fig. \ref{fig_setup_p}d-f), we change the sign of $E_{2\omega}$ in every point in space (Fig. \ref{fig_setup_p}b-e), and thus the field's local handedness (Fig. \ref{fig_setup_p}c-f).
As a result, the dipole of chirality flips direction, and we find the opposite distribution of handedness: ... LR LR LR ...

Polarization of chirality be quantified using chiral correlation functions in the reciprocal space, as described in \cite{Ayuso2021NatComm}.
For the locally chiral field considered in this work, we can use a definition of polarization of chirality along $x$ that is analogous to the polarization of charge\cite{Ayuso2021NatComm},  
\begin{equation}\label{eq_hx}
h_x = \int h^{(5)}(x) x dx
\end{equation}
The amplitude of polarization of chirality maximizes when $\phi_1=\phi_2$, when we have exactly the same amount of left- and right-handed field and the overall structure is achiral ($h_0=0$), as shown in Fig. \ref{fig_setup_p}.

\subsection*{Global chirality versus polarization of chirality}

We have shown how to create synthetic chiral with with different properties: synthetic chiral light which is globally chiral\cite{Ayuso2019NatPhot} and chirality unpolarized (Fig. \ref{fig_setup_g}) and synthetic chiral light that is globally achiral but chirality polarized\cite{Ayuso2021NatComm} (Fig. \ref{fig_setup_p}).
We now show how we can also create light that is both globally chiral and chirality polarized, i.e. with $h_0\neq 0$ and $h_x\neq 0$, although in these cases the values of these quantities are not maximized.

\begin{figure}[ht]
\centering
\includegraphics[width=\linewidth, keepaspectratio=true]{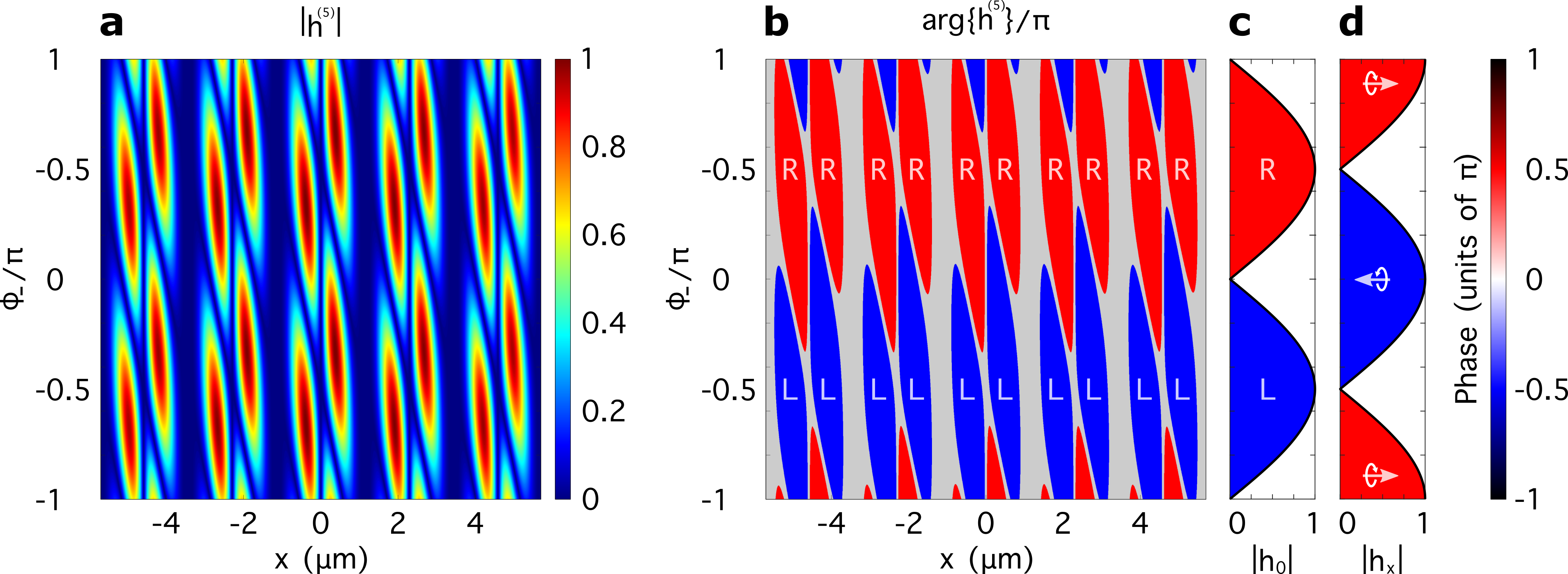}
\caption{\textbf{Local handedness of synthetic chiral light.}
Amplitude (\textbf{a}) and phase (\textbf{b}) of the chiral correlation function $h^{(5)}$, which characterizes the field's local handedness, across the transverse coordinate $x$ as functions of the difference phase delay $\phi_-$, which controls the global chirality properties, for $\phi_+=0$ (see Eq. \ref{eq_phi_pm}).
To calculate the phase of $h^{(5)}$ for a different choice of $\phi_+$, one simply needs to add $\phi_+$ to the values shown in \textbf{b}.
$|h^{(5)}|$ does not depend on $\phi_+$, and thus whether the field is globally chiral and/or chirality polarized depends solely on $\phi_-$.
\textbf{c,} Global chirality $h_0$ (Eq. \ref{eq_h0}) and polarization of chirality $h_x$ (Eq. \ref{eq_hx}).
If $\phi_-=\pm\pi/2$, the field has exactly the same chirality everywhere in space (\textbf{b}), and thus $|h_0|$ maximizes (\textbf{c}) and $h_x=0$ (\textbf{d}).
If $\phi_-=0$ or $\pi$, the field's local handedness changes periodically in space (\textbf{b}) and the overall structure is achiral ($h_0=0$, \textbf{c}) and $|h_x|$ maximizes (\textbf{d}).
For other values of $\phi_-$, the field is both globally chiral and chirality polarized.}
\label{fig_h5}
\end{figure}

Let us analyse the effect of varying $\phi_-$ while keeping $\phi_+$ constant in the spatial distribution of the field handedness.
Fig. \ref{fig_h5} shows the amplitude (Fig. \ref{fig_h5}a) and phase (Fig. \ref{fig_h5}a) of $h^{(5)}$ across the transverse coordinate $x$ as functions of $\phi_-$, for $\phi_+=0$.
This operation moves the $2\omega$-field amplitude grating with respect to the ellipticity grating (see Figs. \ref{fig_setup_g}b,e and \ref{fig_setup_p}b,e) in a way that the shape of the chiral Lissajous figure is not modified other than by changes in the amplitude and sign of the $2\omega$ field component --the two-colour phase delay of the globally chiral field $\phi_+$ is kept constant, see Eq. \ref{eq_E_xt}.
Note that varying $\phi_-$ from $-\pi$ to $\pi$ while keeping $\phi_+=0$ means varying $\phi_1$ from $\pi$ to $-\pi$ and $\phi_2$ from $-\pi$ to $\pi$ synchronously.

For a given value of $\phi_+$, the spatial distribution of $h^{(5)}$ depends on the relative position between the $E_{2\omega}$ and $\varepsilon$ gratings, but the phase of $h^{(5)}$ remains constant up $\pi$ jumps.
If $\phi_-=\pm\pi/2$ ($\phi_2=\phi_1\mp\pi$), these gratings are perfectly aligned (Fig. \ref{fig_setup_g}), and thus the phase $h^{(5)}$ remains constant: the field's chirality is maintained globally in space.
Otherwise, if $\phi_2\neq\phi_1\pm\pi$, the field reverses its handedness periodically in space, and thus the phase of $h^{(5)}$ exhibits $\pi$ jumps along $x$.
As a result, $|h_0|$ maximizes for $\phi_-=\pm\pi/2$, as shown in Fig. \ref{fig_h5}c.
For $\phi_1=\phi_2$ ($\phi_-=0$ or $\pi$), we find the same amount of left- and right-handed field in space, and thus $h_0=0$.

Fig. \ref{fig_h5}d shows the polarization of chirality $h_x$, also as a function of $\phi_-$, for $\phi_+=0$.
For $\phi_-=0$ or $\pi$, where the field is not globally chiral ($|h_0|=0$), the amplitude of polarization of chirality maximizes.
Importantly, the phase of $h_x$ captures the direction of polarization of chirality, which has opposite direction for $\phi_-=0$, where we have the ... RL RL RL ... distribution of handedness, and for $\phi_-=\pi$, where the distribution of handedness is ... LR LR LR ...

We note that, for a given choice of two-colour phase delays $\phi_1$ and $\phi_2$, the definition of $\phi_+$ and $\phi_-$ is not unique.
Let us consider, for instance, the case of $\phi_1=0$ and $\phi_2=\pi$, which leads to $\phi_-=\phi_+=\pi/2$, and thus the field is globally chiral and unpolarized.
The handedness of this field, characterized by the phase of $h^{(5)}$, can be obtained by taking the value of Fig. \ref{fig_h5}b, which assumes $\phi_+=0$, and then adding $\phi_+$ to this value.
For $\phi_-=\pi/2$, the value in Fig. \ref{fig_h5}b is $0.5\pi$ everywhere in space.
Adding $\phi_+$ to this value, we obtain that, for this field, $\arg(h^{(5)})=\pi$ everywhere in space, and thus $\arg(h_0)=\pi$.
Since phase definitions are arbitrary up to $\pm 2n\pi$ ($n\in \mathbb{Z}$), we could represent the exact same field using, e.g. $\phi_1=2\pi$ (instead of $\phi_1=0$) and keeping $\phi_2=\pi$.
In this case, the definition of the sum- and difference-phase delays (Eq. \ref{eq_phi_pm}) changes to $\phi_-=-\pi/2$ and $\phi_+=1.5\pi$, and thus value that we take from Fig. \ref{fig_h5}b is different: $\arg(h^{(5)})=-0.5\pi$.
However, when we add $\phi_+=1.5\pi$ to this value, we obtain the same result: $\arg(h^{(5)})=\pi$ everywhere in space, and thus $\arg(h_0)=\pi$.
Despite this ambiguity in the definition of $\phi_-$ and $\phi_+$, characterizing the field's chirality in terms of these quantities can be advantageous because, as above described, $\phi_-$ unequivocally determines whether the field is globally chiral and/or chirality polarized or not, while $\phi_+$ controls the shape of the chiral Lissajous figure everywhere in space.

\subsection*{Full control over the spatial distribution of handedness}

\begin{figure}[ht!]
\centering
\includegraphics[width=\linewidth, keepaspectratio=true]{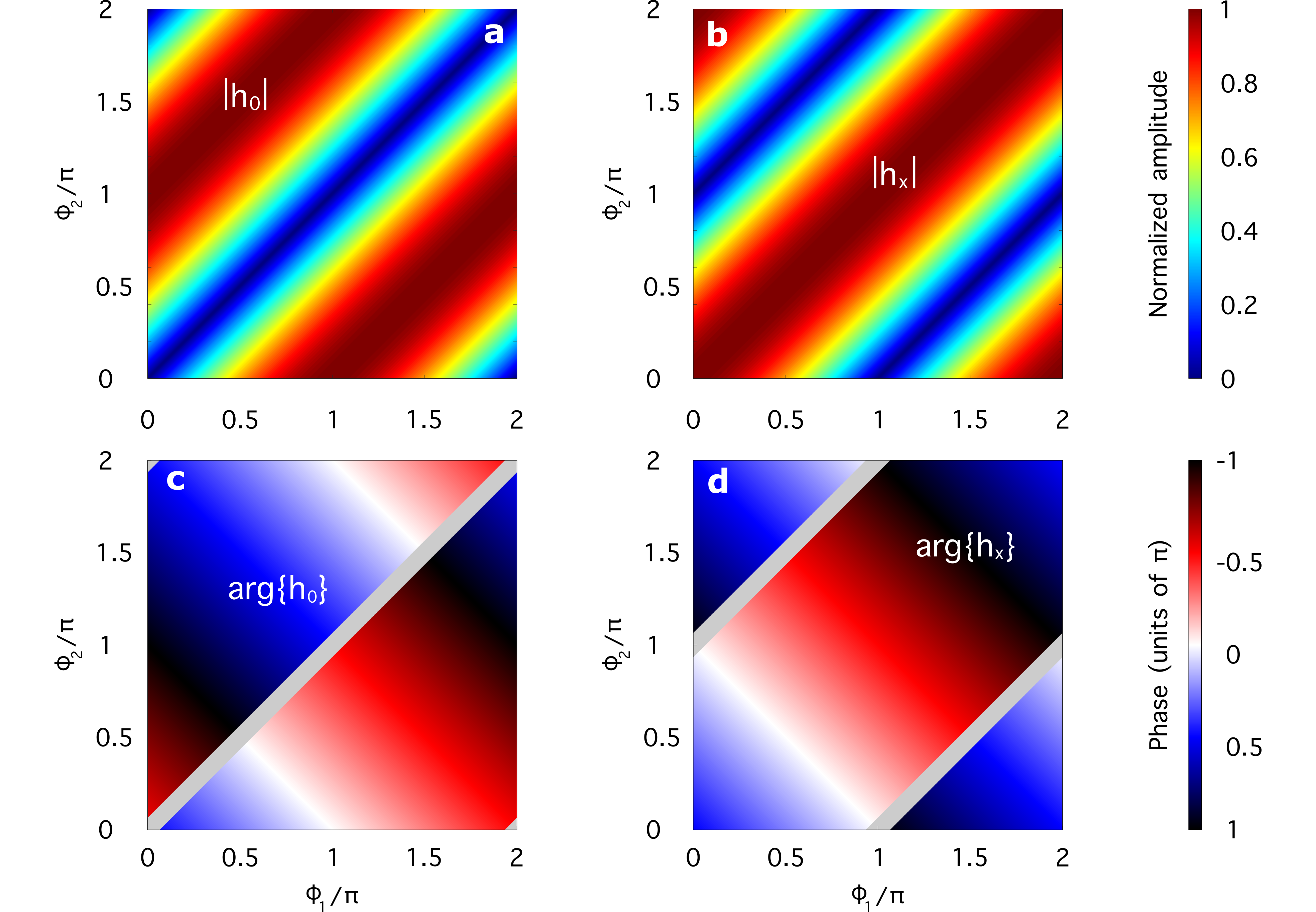}
\caption{\textbf{Global handedness and polarization of chirality.}
Amplitude (\textbf{a}, \textbf{b}) and phase (\textbf{c}, \textbf{d}) of  $h_0$ (\textbf{a}, \textbf{c}, see Eq. \ref{eq_h0}), which characterizes the field's global chirality, and of $h_x$ (\textbf{b}, \textbf{d}, see Eq. \ref{eq_hx}), which characterizes the polarization of chirality, as functions of the two-colour phase delays in the individual beams $\phi_1$ and $\phi_2$ (see Eq. \ref{eq_En}).}
\label{fig_h0x}
\end{figure}

We now explicitly show how the global handedness $h_0$ (Eq. \ref{eq_h0}) and the polarization of chirality $h_x$ (Eq. \ref{eq_hx}) of synthetic chiral light can be fully controlled, both their amplitude and phase, by controlling the two-colour phase delays in our optical setup.
Fig. \ref{fig_h0x} shows how $h_0$ and $h_x$ depend on the individual phase delays $\phi_1$ and $\phi_2$ (see Eq. \ref{eq_En}).
As above discussed, $|h_0|$ maximizes for $\phi_2=\phi_1\pm\pi$, where $|h_x|=0$, and vanishes when $\phi_2=\phi_1$, where $|h_x|$ maximizes.
Both $|h_0|$ and $|h_x|$ remain constant along lines with slope 1, as these lines connect points in which $\phi_2-\phi_1$ remains constant, i.e. lines with the same $\phi_-$ and different $\phi_+$.
The phase of $h_0$ and $h_x$ changes linearly along these lines, reflecting the fact that we tailor the field's local handedness at every point in space, without changing whether the overall structure is global chiral and/or chirality polarized, by varying $\phi_+$.
On the other hand, the lines with slope -1 connect points with constant $\phi_+$ and varying $\phi_-$, and thus the amplitude of both $h_0$ and $h_x$ changes along these lines, while their phase remains constant apart from the $\pi$ jumps associated with the change of sign of $E_{2\omega}$ (see Eq. \ref{eq_E2w}).
This reflects the fact that we can control the global chirality properties of the field (global chirality and polarization of chirality) without changing the shape of the chiral Lissajous figure that the tip of the electric-field vector draws in time, by varying $\phi_-$.
In the next section, we analyze how the local and global properties of synthetic chiral light are imprinted in the macroscopic nonlinear response of randomly oriented chiral molecules.

\section*{Controlling the enantio-sensitive response of chiral molecules}

We have computed the ultrafast electronic nonlinear response of the prototypical chiral molecule H$_2$O$_2$ to synthetic chiral light with different local and global properties.
Here we show how the field properties control the enantio-sensitive response of the medium, and how the molecular handedness can be imprinted into enantio-sensitive macroscopic observables: the total intensity of HHG and the direction of harmonic emission.
The potential energy surface of the electronic ground state of H$_2$O$_2$ has two minima at dihedral angles $112.5^\circ$ and $247.5^\circ$, which correspond to the nuclear configurations of two molecular mirror twins\cite{Kuenzer2019CPL}.
However, because the interconversion energy barrier between them is small, the molecule exists as a racemic mixture in standard conditions.
Still, the relatively small number of nuclei (4) and electrons (18) makes of H$_2$O$_2$ a convenient ``toy model'' to investigate chiral light-matter interactions using reasonable computational resources \cite{Ayuso2022PCCP}.
Here we label the molecular enantiomers with dihedral angles $112.5^\circ$ and $247.5^\circ$ left- and right-handed, respectively.

The single-molecule response of H$_2$O$_2$ to synthetic chiral light has been evaluated using the state-of-the-art implementation of real-time density functional theory in Octopus \cite{Tancogne2020,Andrade2015,Castro2006,Marques2003}, as described in \cite{Ayuso2022PCCP}, running calculations for $200$ different molecular orientations to describe the physical situation of randomly oriented molecules and assuming that the phase of the achiral contribution to light-induced polarization depends linearly on the phase of the $2\omega$ component of the locally chiral field (see Fig. \ref{fig_diagram}b).
We have used a fundamental wavelength $\lambda=400$nm, peak intensity in each beam $I_{\omega}=2.5\cdot10^{13}$ W$\cdot$cm$^{-2}$, beam waist $w=50\mu$m, non-collinear angle $\alpha=10^{\circ}$, and a sine-squared flat-top envelope of 8 laser cycles of the fundamental frequency, with 2 cycles to rise up, 4 cycles of constant intensity, and 2 cycles to go down.
The total peak intensity in the overlap region reaches $10^{14}$ W$\cdot$cm$^{-2}$. 
The amplitude of the $2\omega$ field component was adjusted so the achiral multiphoton pathway (Fig. \ref{fig_diagram}b) giving rise to HHG at frequency $6\omega$ has exactly the same amplitude as the chiral pathway (Fig. \ref{fig_diagram}a) at the same frequency, in order to maximize the enantio-sensitive interference between the two contributions.
Our numerical simulations show that this condition is achieved by setting $\sqrt{I_{2\omega}/I_{\omega}}=0.0125$ in each beam.

\subsection*{Angularly resolved harmonic spectra}

\begin{figure}[ht!]
\centering
\includegraphics[width=\linewidth, keepaspectratio=true]{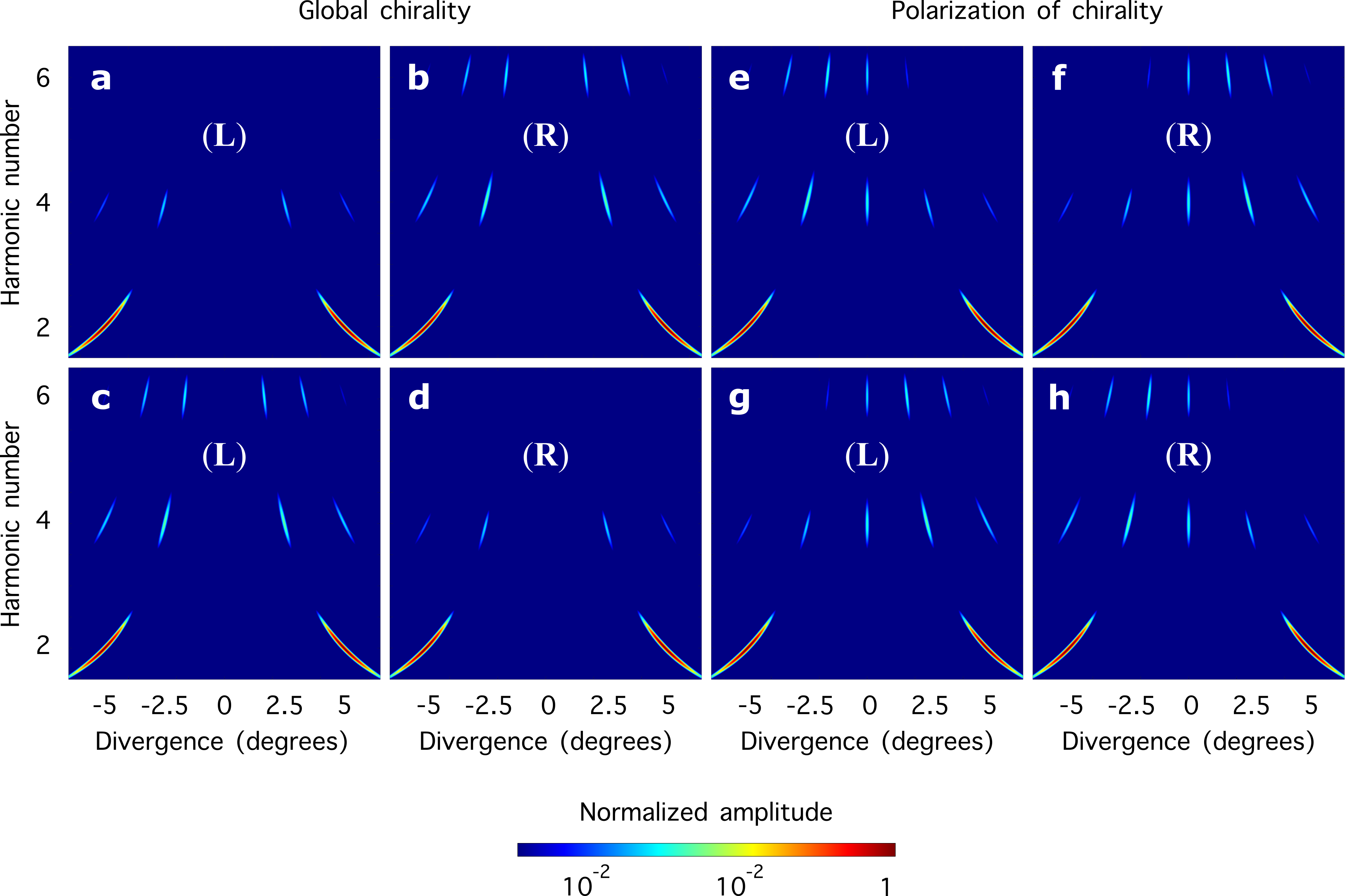}
\caption{\textbf{Enantio-sensitive response of chiral matter to synthetic chiral light in the far field.}
Amplitude of the enantio-sensitive $z$-polarized component of the harmonic light emitted from left- and right-handed H$_2$O$_2$ upon interaction with synthetic chiral light that is globally chiral and chirality unpolarized (\textbf{a}-\textbf{d}) and globally achiral and chirality polarized (\textbf{e}-\textbf{h}) as a function of the harmonic frequency and the divergence angle.
To maximize $|h_0|$ and suppress $|h_x|$ (\textbf{a}-\textbf{d}), we impose $\phi_2=\phi_1+\pi$ ($\phi_-=\pi/2$).
Setting $\phi_1=0.17\pi$ and $\phi_2=1.17\pi$ ($\phi_+=0.67\pi$) quenches emission at frequency $6\omega$ in the left-handed molecule (\textbf{a}) while maximizing it in the right-handed molecule (\textbf{b}).
Changing $\phi_1$ and $\phi_2$ by $\pi$ ($\phi_1=1.17\pi$ and $\phi_2=0.17\pi$) changes the field's local handedness in every point in space, leading to enhancement of emission at frequency $6\omega$ if the left-handed molecule (\textbf{c}) and suppression in the right-handed molecule (\textbf{d}). 
To suppress $|h_0|$ and maximize $|h_x|$ (\textbf{e}-\textbf{h}), we impose $\phi_1=\phi_2$ ($\phi_-=0$).
Setting $\phi_1=\phi_2=1.17\pi$ ($\phi_+=1.17\pi$) bends emission of harmonic 6 to the left in the left-handed molecule (\textbf{e}) and to the right (\textbf{f}) in the right-handed molecule.
Changing $\phi_1$ and $\phi_2$ by $\pi$ ($\phi_1=\phi_2=0.17\pi$) changes the field's local handedness in every point in space, reversing the direction of polarization of chirality and thus the direction of enantio-sensitive light bending (\textbf{g},\textbf{h}).}
\label{fig_ff}
\end{figure}

The amplitude of harmonic light emitted from the left- and right-handed enantiomers of H$_2$O$_2$ upon interaction with synthetic chiral light that is and globally chiral and unpolarized is presented in Fig. \ref{fig_ff}a-d, as a function of the harmonic frequency and the divergence angle.
We show the enantio-sensitive $z$-polarized component of the emitted harmonic light, which contains even harmonic frequencies.
The $x$-polarized component (not shown) is not enantio-sensitive, but it is separated from the enantio-sensitive component in frequency, polarization and space\cite{Ayuso2019NatPhot}.
To maximize the global chirality $|h_0|$, we set $\phi_-=\pi/2$, i.e. $\phi_2=\phi_1+\pi$.
As above described, in this situation $E_{2\omega}$ and $\varepsilon$ flip sign exactly at the same positions, see Fig. \ref{fig_setup_g}b,e, and the field is globally chiral and chirality unpolarized ($h_x=0$), see Figs. \ref{fig_h5} and \ref{fig_h0x}.
We have adjusted $\phi_+$ so the two contributions to light-induced polarization at frequency $6\omega$ are perfectly out of phase in the left-handed molecules (Fig. \ref{fig_ff}a), and thus perfectly in phase in the right-handed molecules (Fig. \ref{fig_ff}b).
Our simulations reveal that, for this harmonic frequency, this condition is achieved by setting $\phi_+=0.67\pi$, i.e. by setting $\phi_1=0.17\pi$ and $\phi_2=1.17\pi$.
If we change the phase delay in the two beams by $\pi$ ($\phi_1=1.17\pi$ and $\phi_2=0.17\pi$), we reverse the field's local handedness in each point in space while keeping the overall structure globally chiral and unpolarized, and then we obtain the opposite result: we enhance harmonic generation from the left-handed enantiomer (Fig. \ref{fig_ff}c) and quench it in the right-handed enantiomer (Fig. \ref{fig_ff}d).
That is, by adjusting $\phi_1$ and $\phi_2$ in a way that we keep $\phi_2=\phi_1\pm\pi$, we can adjust the handedness of the field everywhere in space to maximize enantio-sensitive response of the medium at the level of total intensity signals.

The macroscopic response of the molecules to synthetic chiral light that is globally achiral, but chirality polarized, is presented in Fig. \ref{fig_ff}e-h.
Here, we maximize $|h_x|$ by setting $\phi_1=\phi_2$, so that $\phi_-=0$, and thus $E_{2\omega}$ and $\varepsilon$ change sign along $x$ with a spatial phase delay of $\pi/2$, as shown in Fig. \ref{fig_setup_p}b,e.
The total intensity of emission is not enantio-sensitive because now the field is globally achiral ($h_0=0$).
However, because the field is chirality polarized ($h_x\neq0$), the molecular handedness is imprinted in the direction of harmonic emission.
Setting $\phi_1=\phi_2=1.17\pi$ (and thus $\phi_+=1.17\pi$) bends the nonlinear response of the left- and right-handed molecules to the left and to the right, respectively.
If we set $\phi_1=\phi_2=0.17\pi$ instead, then we reverse the polarization of chirality, and now the left-handed molecules emit light at frequency $6\omega$ to the right, while the right-handed molecules emit light to the left.
Thus, by adjusting $\phi_1$ and $\phi_2$ simultaneously, while keeping $\phi_1=\phi_2$, we can achieve full control over the enantio-sensitive direction of light bending. 

Note that the value of $\phi_+$ that maximizes the enantio-sensitive response of the molecules to synthetic chiral that is globally chiral ($h_0\neq0$) and unpolarized  ($h_x=0$) is different from the value that maximizes the enantio-sensitive response to light that is globally achiral ($h_0=0$) and polarized ($h_x\neq0$).
As above described (see Fig. \ref{fig_diagram}), these enantio-sensitive signals rely on the interference between two contributions to light-induced polarization at the same frequency: a chiral contribution $P_c$ and an achiral contribution $P_a$.
To maximize the enantio-sensitive response at the level of total intensity signals when the field is globally chiral and unpolarized, we need to adjust $\phi_+$ so that $P_a$ and $P_c$ are exactly in phase in one enantiomer (and out of phase in the other) at each point in space, in the near field.
As a result, the intensity of the optical response is enantio-sensitive already at the single-molecule response level.
The mechanism giving rise to enantio-sensitive light bending when the field is chirality polarized is different.
It relies on the interference between the radiation emitted from $P_a$ and $P_c$ in the far field.
This mechanism is optimized when $P_a$ and $P_c$ radiate with a phase delay of $\pm\pi/2$.
That is, in this case, the two contributions to light-induced polarization do not interfere at the single-molecule response level, but in the macroscopic harmonic emission.
As a result, the values of $\phi_+$ that maximize the enantio-sensitive intensity of harmonic emission when light is globally chiral and unpolarized are shifted by $\pi/2$ with respect to the values that maximize enantio-sensitive light bending when light is globally achiral and polarized.

\subsection*{Control over the total intensity of harmonic emission}

\begin{figure}[ht]
\centering
\includegraphics[width=\linewidth, keepaspectratio=true]{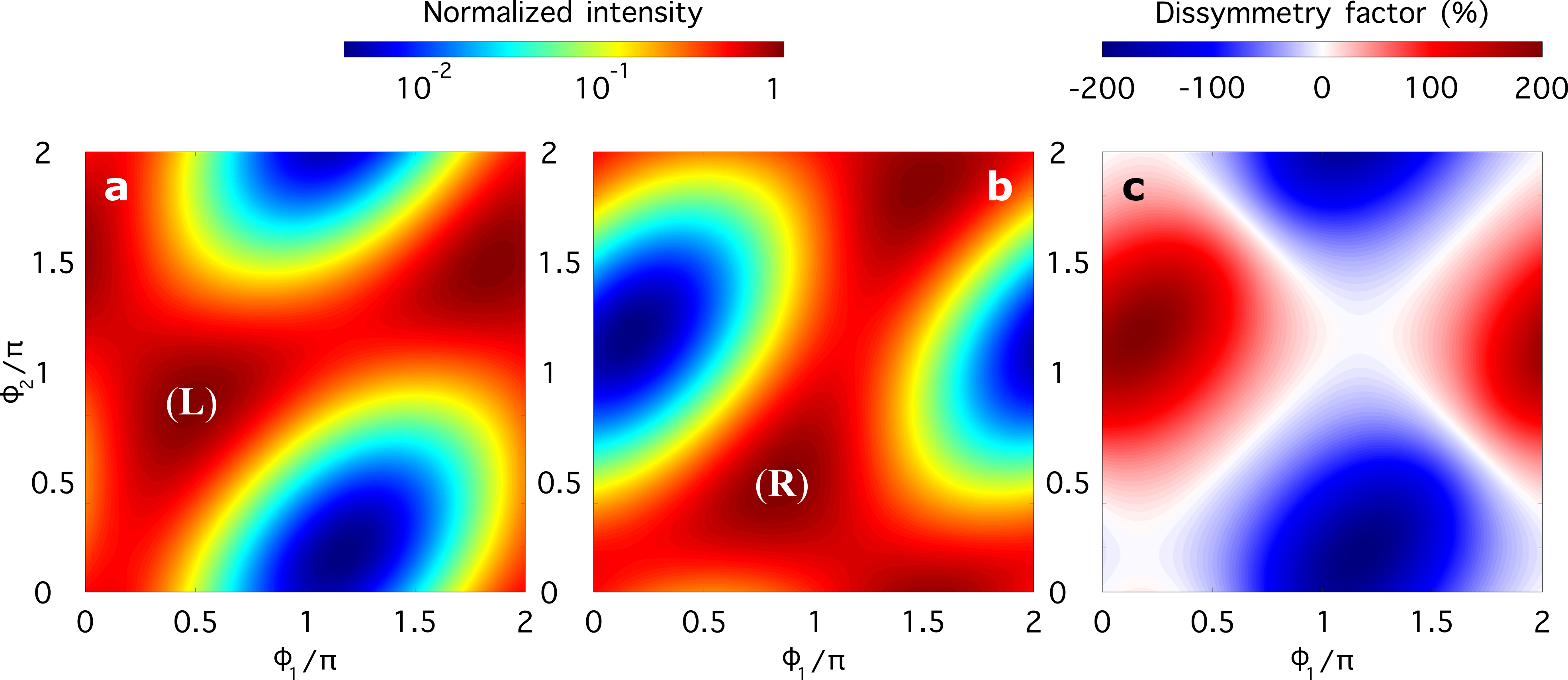}
\caption{\textbf{Control over the total intensity of harmonic emission.}
Total intensity of harmonic light, integrated over the divergence angle, emitted from left- (\textbf{a}) and right-handed (\textbf{b}) H$_2$O$_2$ at frequency $6\omega$, and dissymmetry factor $\gamma$ (\textbf{c}, see Eq. \ref{eq_gamma}) as functions of the two-colour phase delays in the two beams.
The enantio-sensitive response maximizes for $\phi_2=\phi_1+\pi$ because in this case the field's global handedness, quantified by $h_0$ (Eq. \ref{eq_h0}), maximizes.
For $\phi_2=\phi_1$, the field is not globally chiral ($h_0=0$), and thus the total intensity of harmonic emission is not enantio-sensitive.
Note that, in this case, the enantio-sensitive observable is the direction of harmonic emission, see Fig. \ref{fig_angle}.}
\label{fig_I+CD}
\end{figure}

We now show how the total intensity of harmonic emission, integrated over the divergence angle, can be controlled by controlling the individual two-colour phase delays $\phi_1$ and $\phi_2$ (Eq. \ref{eq_En}).
Fig. \ref{fig_I+CD} shows the total intensity of harmonic emission from the left-handed ($I_L$, Fig. \ref{fig_I+CD}a) and right-handed ($I_R$, Fig. \ref{fig_I+CD}b) enantiomers at frequency $6\omega$, as functions of $\phi_1$ and $\phi_2$.
To quantify the enantio-sensitive response, we use a standard definition of the dissymmetry factor,
\begin{equation}\label{eq_gamma}
\gamma=2\frac{I_L-I_R}{I_L+I_R}.
\end{equation}
The values of $\gamma$ are presented in Fig. \ref{fig_I+CD}c, also as a function of $\phi_1$ and $\phi_2$.
These phase scans record the global chirality properties of synthetic chiral light and provide valuable information about the ultrafast electronic response of the chiral molecules.
If $\phi_1=\phi_2$, then the field is not globally chiral ($h_0=0$, see Fig. \ref{fig_h0x}a) and thus the total intensity of harmonic emission cannot be enantio-sensitive, $I_L=I_R$.
This is the origin of the white ($\gamma=0$) line with slope 1 in Fig. \ref{fig_I+CD}c, which is characteristic of the field configuration and will be present when recording the enantio-sensitive response of any chiral molecule, and at any (even) harmonic frequency.
The white line with slope -1 in Fig. \ref{fig_I+CD}c, however, is a molecule-specific feature.
It reflects the fact that the enantio-sensitive response of the molecule maximizes when the phase delay between the achiral $P_a$ and chiral $P_c$ contributions to light-induced polarization is $0$ or $\pi$, so they interfere constructively in one enantiomer and destructively in the other, already at the single-molecule response level.
If the phase delay between $P_a$ and $P_c$ is $\pm\pi/2$, they do not interfere in the near field, and thus the total intensity emitted from left- and right-handed molecules is identical, yielding $\gamma=0$.
Note that $\gamma=0$ does not mean that the response of the chiral molecules is not enantio-sensitive.
While the total intensity signal emitted from opposite enantiomers is the same when the field is not globally chiral ($h_0=0$), for these field configurations the field has polarization of chirality ($h_x\neq0$), which leads to enantio-sensitive light bending.
As shown in Fig. \ref{fig_I+CD}, by adjusting $\phi_1$ and $\phi_2$, we fully control the total intensity of harmonic emission, and the dissymmetry factor reaches the ultimate limits of $\pm200\%$.

\subsection*{Control over enantio-sensitive light bending}

\begin{figure}[ht]
\centering
\includegraphics[width=\linewidth, keepaspectratio=true]{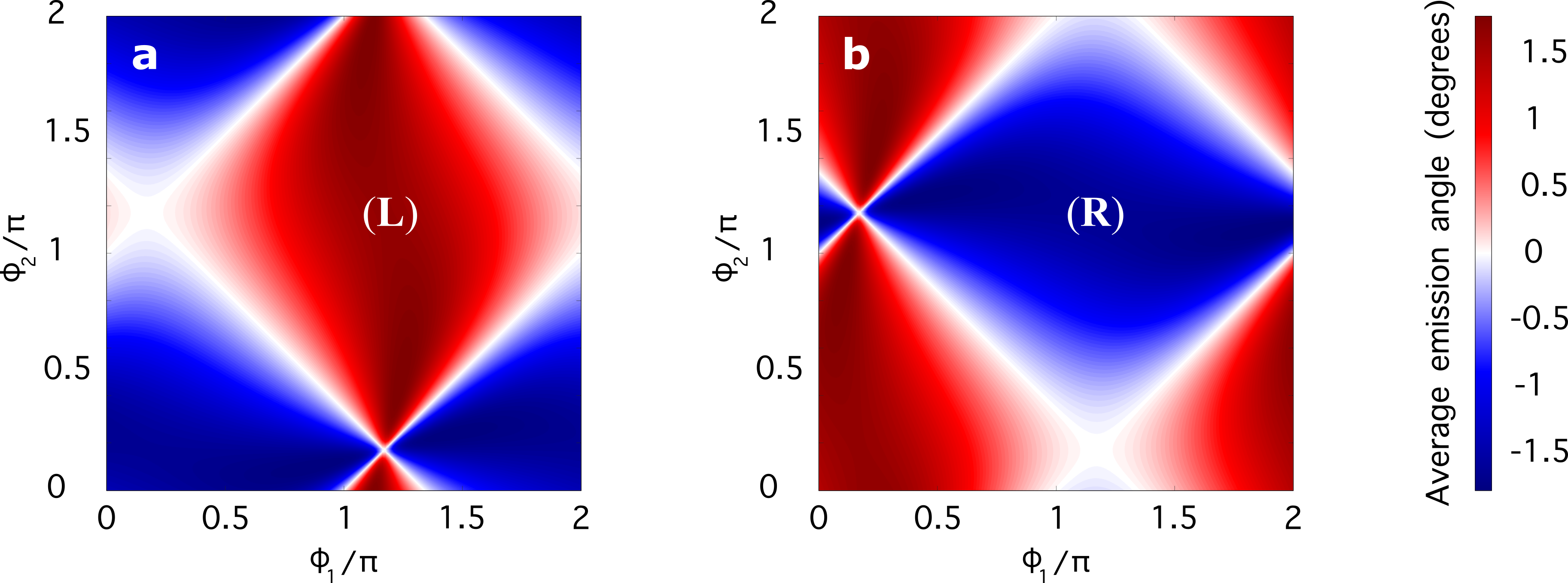}
\caption{\textbf{Control over the direction of harmonic emission.}
Average angle of harmonic emission (Eq. \ref{eq_beta}) at frequency $6\omega$ from left-handed (\textbf{a}) and right-handed (\textbf{b}) H$_2$O$_2$, as a function of the two-colour phase delays in the two beams.
The enantio-sensitive direction of emission maximizes for $\phi_2=\phi_1$ because in this case the field's polarization of chirality, quantified by $h_x$ (Eq. \ref{eq_hx}), maximizes.
For $\phi_2=\phi_1+\pi$, the field is chirality unpolarized ($h_x=0$), and thus the direction of harmonic emission is not enantio-sensitive.
Note that, in this case, the total intensity of harmonic emission is highly enantio-sensitive, see Fig. \ref{fig_I+CD}.}
\label{fig_angle}
\end{figure}

We now explicitly show how the direction of harmonic emission can be controlled by controlling the individual phase delays $\phi_1$ and $\phi_2$ (Eq. \ref{eq_En}).
To quantify the degree of enantio-sensitive light bending, we calculate the average angle of harmonic emission,
\begin{equation}\label{eq_beta}
\langle\beta\rangle = \int \frac{\int I(\beta)\beta d\beta}{\int I(\beta) d\beta},
\end{equation}
where $\beta$ is the divergence angle.
Fig. \ref{fig_angle} shows the average angle of emission at the harmonic frequency $6\omega$ from the left- (Fig. \ref{fig_angle}a) and right-handed (Fig. \ref{fig_angle}b) molecules.
Note that $|\langle\beta\rangle|$ maximizes in the phase regions where the total intensity of harmonic emission, integrated over $\beta$, is not enantio-sensitive ($\gamma=0$, see Fig. \ref{fig_I+CD}), and it is suppressed in the phase regions where $|\gamma|$ maximizes.
Like the $\gamma$ spectrogram presented in Fig. \ref{fig_I+CD}, the scans over $\langle\beta\rangle$ also contain two white lines where $\langle\beta\rangle=0$.
Here, the white line with slope 1 appears at $\phi_2=\phi_1+\pi/2$, where the field is globally chiral ($h_0\neq 0$) and unpolarized ($h_x=0$), see Fig. \ref{fig_h0x}c.
This line is characteristic of the specific field configuration and independent of the molecular properties, like the white line of slope 1 the $\gamma$ spectrogram. 
Also like in the $\gamma$ spectrogram, the white line with slope $-1$ records the ultrafast electronic response of the chiral medium, and it a is molecule-specific feature.
Here, the white line with slope -1 appears in the region where $P_a$ and $P_c$ are exactly in phase in one enantiomer and out of phase in the other,
because enantio-sensitive light bending requires interference in the far field, which is maximized when $P_a$ and $P_c$ have a phase delay of $\pm\pi/2$, and suppressed when they have a phase delay of $0$ or $\pi$.
As shown in Fig. \ref{fig_angle}, by adjusting $\phi_1$ and $\phi_2$, we fully control the direction of enantio-sensitive light bending, and the average emission angles reach high values, up to $\pm1.5^\circ$.

\section*{Discussion}

Synthetic chiral light is a powerful optical tool for chiral recognition, which allows us to bypass two fundamental limitations of (standard) circularly polarized light.
The first limitation is related to the strength of the chiro-optical signals.
Circularly polarized light is not chiral within the electric-dipole approximation: the Lissajous figure that the tip of the electric-field vector draws in time, in a given point in space, is a (planar) circle.
This circle becomes chiral once we include the propagation vector of the light wave and its magnetic-field component, but non-electric-dipole interactions are weak for small to medium-size molecules, and lead to weak enantio-sensitivity.
In contrast, the enantio-sensitive response of chiral matter to synthetic chiral light is solely driven by the electric-field vector of the light field, and is orders of magnitude stronger.

The second constraint of standard chiral light is that it offers limited opportunities for control: we can change between left and right circular or elliptical polarization, but the phase between the electric and magnetic components of the wave is fundamentally locked.
As a result, the enantio-sensitive response of the molecules is not only weak, but also harder to control.
In contrast, the handedness of synthetic chiral light depends on the phase delay between its different frequency components, which is a continuous and controllable parameter.
By adjusting the phase delays in our non-collinear setup, we can shape the Lissajous figure that the tip of the electric-field vector draws in time, in every point in space, in order to maximize the enantio-sensitive response of the chiral medium.

Here we have shown how to spatially structure the local handedness of synthetic chiral light to create optical fields with different local and global properties.
Using state-of-the-art computational modelling to evaluate the ultrafast electronic response of randomly oriented chiral molecules, we have analyzed how these local and global properties are imprinted into enantio-sensitive macroscopic observables.
On the one hand, synthetic chiral light that is locally and globally chiral can enhance the nonlinear response of a selected molecular enantiomer while quenching it in its mirror twin at the level of total intensity signals.
On the other hand, synthetic chiral light that exhibits polarization of chirality can bend the nonlinear response of chiral matter in an enantio-sensitive manner, forcing the left-handed molecules to emit light, e.g. to the left, while the right-handed molecules emit light to the right.
In contrast with our previous works, we have shown how one can engineer synthetic chiral light that is both globally chiral and chirality polarized.
Here, the intensity of harmonic emission is both enantio-sensitive at the level of total intensity signals, and sent in different directions from opposite molecular enantiomers, although the degree of enantio-sensitivity in both macroscopic observables is reduced.

Our analysis reveals the presence of enantio-sensitive molecule-specific fingerprints in the spectra of harmonic emission.
Indeed, the nodal lines in the 2D spectrograms which record the dissymmetry factor (Fig. \ref{fig_I+CD}) and the average emission angles (Fig. \ref{fig_angle}) as functions of the two-colour phase delays in the incident beams, contain valuable information of the laser field configuration, which could be used for calibration purposes, but also of enantio-sensitive interplay between the chiral and achiral components of the ultrafast dynamics induced in the molecules.

The possibility of creating synthetic chiral light with controllable local handedness, controllable global handedness, and controllable polarization of chirality, creates new avenues for efficient chiral recognition on ultrafast time scales.
New opportunities may arise from the possibility of sculpting the local handedness of synthetic chiral light using vortex beams, taking advantage of the structured phase profiles characteristic of such waveforms.

\section*{Acknowledgments}
D. A. acknowledges endless support, guidance and inspiration from Olga Smirnova and Misha Ivanov, and enlightening discussions with them and with Andr\'es Ordoñez.
L. R acknowledges funding from the European Union-NextGenerationEU and the Spanish Ministry of Universities via her Margarita Salas Fellowship through the University of Salamanca;
D. A. acknowledges funding from the Deutsche Forschungsgemeinschaft SPP 1840 SM 292/5-2;
L. R. and D. A. acknowledge funding from the Royal Society URF$\backslash$R1$\backslash$201333 and RF$\backslash$ERE$\backslash$210358.

\bibliography{Bibliography}

\end{document}